\documentclass{desyproc}

\def\beq{\begin{equation}}
\def\eeq{\end{equation}}
\def\etal{{\it et al.}}
\def\ga{\mathrel{\raise.3ex\hbox{$>$\kern-.75em\lower1ex\hbox{$\sim$}}}}
\def\la{\mathrel{\raise.3ex\hbox{$<$\kern-.75em\lower1ex\hbox{$\sim$}}}}
\newcommand{\gev}{\,\, \mathrm{GeV}}
\newcommand{\tb}{\tan\beta}
\newcommand{\Mh}{M_h}

\newcommand{\fNTq}[1]{\ensuremath{f_{T_{#1}}^{(N)}}}

\newcommand{\BNq}[1]{\ensuremath{B_{#1}^{(N)}}}

\newcommand{\SigmapiN}{\ensuremath{\Sigma_{\pi\!{\scriptscriptstyle N}}}}
\newcommand{\ssi}{\sigma^{\rm SI}_p}
\newcommand{\neu}[1]{\tilde \chi^0_{#1}}
\newcommand{\mneu}[1]{m_{\tilde \chi^0_{#1}}}

\begin{document}
\title{Dark Energy and Dark Matter}

\author{{\slshape Keith A. Olive$^1$}\\[1ex]
$^1$William I. Fine Theoretical Physics Institute, University of Minnesota, Minneapolis, MN 55455
}

\contribID{35}

\confID{800}  
\desyproc{DESY-PROC-2009-xx}
\acronym{LP09} 
\doi  

\maketitle

\begin{abstract}
\vskip -2.5in
{\rightline{\small UMN--TH--2836/10, FTPI--MINN--10/05}}
\vskip 2.3in
A brief overview of our current understanding of abundance and properties of
dark energy and dark matter is presented.  A more focused discussion of supersymmetric dark 
matter follows. Included is a frequentist approach to the supersymmetric parameter space
and consequences for the direct detection of dark matter.
\end{abstract}

\section{The Energy Density Content of the Universe}

The overall composition of the Universe can be conveniently described by the density parameter,
$\Omega$, defined as the average energy density of the Universe, $\rho$, relative to the critical density
needed for a spatially flat Universe, $\rho_c$.  One of the Einstein field equations
leads to the expression for the expansion rate of the Universe, which we characterize by
the Hubble parameter,
\beq
H^2  \equiv \left({\dot{R} \over R}\right)^2  = { 8 \pi G_N \rho \over 3}
 - { k \over R^2}  + {\Lambda \over 3} ,
\label{H}
\eeq
where $R(t)$ is the cosmological scale factor and $k$ is the three-space
curvature constant ($k = 0, +1, -1$ for a spatially flat, closed or open
universe). $\Lambda$ is the cosmological constant which is assumed here to contain
all contributions from the vacuum energy density.
One can 
define a critical energy density $\rho_c$
  such that $\rho =\rho_c$  for $k = 0$
\beq
	\rho_c  = 3H^2 / 8 \pi G_N		.
\eeq
In terms of the present value of the Hubble parameter this is,
\beq
	\rho_c  = 1.88 \times 10^{-29} {h_0}^2  {\rm g cm}^{-3}  ,
\eeq
where
\beq
	h_0  = H_0 /(100 {\rm km Mpc}^{-1}   {\rm s}^{-1}  )		.
\eeq
The cosmological density parameter is then defined by
\beq
	\Omega \equiv {\rho \over \rho_c} 	.		
\eeq

The composition of the Universe can be expressed by breaking down
the density parameter into separate contributions,
\beq
\Omega = \Omega_r + \Omega_m + \Omega_\Lambda ,
\eeq
for contributions from radiation, matter and a cosmological constant/vacuum
with $\Omega_\Lambda = \Lambda/3H^2$.
The contribution to $\Omega_r$ from the cosmic microwave background (CMB) is small,
of order $10^{-4}$.  Precise determinations of the matter and vacuum
contributions to $\Omega$ are obtained from the detailed power spectrum of CMB anisotropies
as measured by WMAP \cite{wmap}.  When combined with other measurements such as
high redshift  supernova type Ia data \cite{sn1} and baryon acoustic oscillations \cite{bao}, one finds
\beq
h_0 = 0.71 \pm 0.01  \qquad \Omega_0 = 1.006 \pm 0.006.
\label{range}
\eeq
WMAP data alone is sufficient for determining the individual contributions to $\Omega$ 
of
\beq
\Omega_m h_0^2 = 0.133 \pm 0.006 \qquad \Omega_\Lambda = 0.74 \pm 0.03 .
\label{oh2}
\eeq

The matter content of the Universe can be further broken down as WMAP
also determines the baryon density of Universe \cite{wmap}
\beq
\Omega_B h_0^2 = 0.0227 \pm 0.0006 .
\eeq
The contribution to $\Omega$ in neutrinos lies in the range
\beq
0.0005 < \Omega_\nu h_0^2 < 0.0076 ,
\eeq
where the lower bound is obtained from the requirement of finite neutrino masses
from oscillation data and the upper bound is again derived from WMAP data
in conjunction with other large scale structure data.

\section{Dark Energy}

The biggest surprise of all of the recent determinations of contributions to $\Omega$ must be
the realization that there is a substantial contribution from dark energy, namely that
$\Omega_\Lambda \ne 0$. The WMAP value for $\Omega_\Lambda$
is moreover consistent with determinations from supernovae data and baryon acoustic oscillations.
When all data are used, one finds $\Omega_\Lambda = 0.726 \pm 0.015$. 

But now a bigger question arises:  What is the physical nature of the dark energy?
Different possibilities can be distinguished by their equation of state characterized by
$w = p/\rho$.  The equation of state parameter for radiation is simply $w_r = 1/3$, whereas for
matter, it is $w_m = 0$.

The simplest solution for the dark energy remains either a cosmological constant
or a constant vacuum contribution to the energy density with an equation of state, $w = -1$. 
This is indeed consistent with the central value determined by WMAP, which finds \cite{wmap}
\beq
-0.33 < 1 + w_0 < 0.21 ,
\eeq
for the value of $w$ today (at 95 \% CL).
The numerical value for $\Lambda$, however, is extremely small, and when
written as a dimensionless constant (as $G_N \Lambda$), it is of order $10^{-123}$.
This is the well known cosmological constant problem in cosmology \cite{cc}.

There are, of course, other possibilities, the largest class of which is known as quintessence \cite{quint}.
In this case, the dark energy may be a dynamical phenomenon described by an evolving
scalar field. The energy density and pressure of a scalar field, $\phi$, with potential, $V(\phi)$,
can be written as (neglecting spatial gradient terms)
\begin{eqnarray}
	\rho =  {1 \over 2} {\dot{\phi}}^2  + V(\phi)		\\
	p =  {1 \over 2} {\dot{\phi}}^2  -  V(\phi)	.
\label{prho}
\end{eqnarray}
When the kinetic term is small compared to the potential, $\rho \approx V$ and
$p \approx -V$, and we recover the constant solution with $w = -1$. 
In general, however, $w_0$ may differ from -1 and indeed may not even be a constant.
Once again, WMAP (using supernovae and BAO data) place combined limits on 
$w$ and its derivative with respect to redshift, $w^\prime$,
\beq
w = -1.06 \pm 0.14 \qquad w^\prime = 0.36 \pm 0.62
\eeq

In short summary, we are left with the following puzzles regarding dark energy:
\begin{itemize}
\item There is the question of fine-tuning associated with the cosmological constant problem.
Namely, we expect several contributions to the vacuum energy density 
\beq
\Lambda = \Lambda_{GUT} + \Lambda_{EW} + \Lambda_{QCD} \cdots
\eeq
where the various contributions listed arise from possible sources such as 
grand unified theories ($ G_N \Lambda_{GUT} \sim (10^{-3})^{4}$) ,
the Standard Model ($ G_N \Lambda_{EW} \sim (10^{-16})^{4}$), 
and QCD ($ G_N \Lambda_{QCD} \sim (10^{-20})^{4}$), yet sum to $10^{-123}$.
\item The coincidence problem. Here, we would like to understand why 
$\Omega_m$ is within a factor of a few of $\Omega_\Lambda$ today.
This is one of the issues addressed by quintessence models and may
be probed in observations testing the possibility of variability in fundamental constants.
\end{itemize}

\section{Dark Matter}

From the quoted contributions to $\Omega$ in matter and baryons from WMAP, we can obtain the density of cold dark matter from
the difference between the total matter density and the baryon density \cite{wmap}
\beq
\Omega_{CDM} h^2 = 0.1099 \pm 0.0062
\label{wmap}
\eeq
or a 2$\sigma$ range of 0.0975 -- 0.1223 for $\Omega_{CDM} h^2$.

Evidence for dark matter in the universe is available from a wide range of
observational data.  In addition to the results from the CMB,
there is the classic evidence from galactic rotation curves \cite{rot}, which indicate
that nearly all spiral galaxies are embedded in a large galactic halo of dark matter
leading to rather constant rotational velocities at large distances from the center of the galaxy
(in contrast to the expected $v^2 \sim 1/r$ behavior in the absence of dark matter).
Other dramatic pieces of evidence can be found in combinations of X-ray observations
and weak lensing showing the superposition of dark matter (from lensing) and ordinary matter
from X-ray gas \cite{witt} and from the separation of baryonic and dark matter
after the collision of two galaxies as seen in the Bullet cluster \cite{clowe}.
For a more complete discussion see \cite{otasi3}.

In addition to being stable (or at least very long lived),
the dark matter should be both electrically and color neutral.
Indeed, there are very strong constraints, forbidding the existence of stable or
long lived particles which are not color and electrically neutral as
these would become bound with normal matter
forming anomalously heavy isotopes. The limits on the
abundances, relative to hydrogen, of nuclear isotopes \cite{isotopes},
$n/n_H \la 10^{-15}~~{\rm to}~~10^{-29}$
for 1 GeV $\la m \la$ 1 TeV. A strongly interacting stable relic is expected
to have an abundance $n/n_H \la 10^{-10}$
with a higher abundance for charged particles. 

Unfortunately, there are no viable candidates for dark matter in the Standard Model.
As baryons and neutrinos have been excluded, one is forced to go beyond the Standard Model,
and here, I will focus on the possibilities which exist in the context of the minimal supersymmetric extension of the Standard Model (MSSM) \cite{mssm}. 
In the MSSM, the lightest supersymmetric particle (LSP) is stable if R-parity ($R = -1^{3B + L +2s}$) 
is unbroken.
There are several possibilities
in the MSSM,
specifically the sneutrino with spin zero, the neutralino with spin 1/2, and the gravitino
with spin 3/2. However, a sneutrino LSP would have relatively large coherent interactions
with heavy nuclei, and experiments searching directly for the scattering of 
massive dark matter particles on nuclei exclude a stable sneutrino weighing
between a few GeV and several TeV \cite{Falk:1994es}. The possible loophole of a very light
sneutrino was excluded by measurements of the invisible $Z$-boson decay rate at LEP
\cite{LEP:2003ih}. The gravitino is a viable candidate and often predicted in models
based on supergravity \cite{vcmssm,gdm}.  In this case, however, its probability for direct detection is
negligible.

There are four neutralinos, each of which is a  
linear combination of the $R=-1$ neutral fermions \cite{EHNOS}: the wino
$\tilde W^3$, the partner of the
 3rd component of the $SU(2)_L$ gauge boson;
 the bino, $\tilde B$;
 and the two neutral Higgsinos,  $\tilde H_1$ and $\tilde H_2$.
The mass and composition of the LSP are determined by the gaugino masses, $M_1$ and $M_2$,
the Higgs mixing mass term, 
$\mu$, and  the ratio of the two Higgs vacuum expectation 
values expressed as $\tan \beta$. In general,
neutralinos can  be expressed as a linear combination
\begin{equation}
	\chi = \alpha \tilde B + \beta \tilde W^3 + \gamma \tilde H_1 +
\delta 
\tilde H_2 .
\end{equation}

The relic density of neutralinos depends on additional parameters in the MSSM beyond $M_1, M_2,
\mu$, and $\tan \beta$. These include the sfermion masses, $m_{\tilde f}$ and the
Higgs pseudo-scalar mass, $m_A$. To
determine the relic density it is necessary to obtain the general
annihilation cross-section for neutralinos.  In much of the parameter
space of interest, the LSP is a bino and the annihilation proceeds mainly
through sfermion exchange.

In its generality, the MSSM has over 100 undetermined
parameters.There are good arguments based on grand unification \cite{GUT} and supergravity \cite{BIM} which lead to a strong reduction in the number of parameters.
I will assume several unification conditions placed on the
supersymmetric parameters.  In all models considered, the gaugino masses
are assumed to be unified at the GUT scale with value, $m_{1/2}$, 
as are the trilinear couplings with value $A_0$.  Also common to all models
considered here is the unification of all soft scalar masses set equal to $m_0$ at the GUT scale.
With this set of boundary conditions at the GUT scale, we can use the radiative electroweak 
symmetry breaking conditions by specifying $\tan \beta$, and the mass, $M_Z$, to predict the 
values of $\mu$ and Higgs pseudoscalar mass, $m_A$.
The sign of $\mu$ remains free. 
This class of models is often referred to as the constrained MSSM (CMSSM) \cite{funnel,cmssm,efgosi,eoss,cmssmwmap}.
In the CMSSM, the solutions for $\mu$ generally lead to a lightest neutralino
which is very nearly a pure $\tilde B$. 

I note that that while the name CMSSM is often used synonymously 
with mSUGRA, for minimal supergravity \cite{BIM,bfs}.  The latter however, has two additional
constraints: $m_{3/2} = m_0$ and $B_0 = A_0 - m_0$.
The former sets the unification scale scalar masses equal to the gravitino mass.
This condition often results in a gravitino LSP \cite{vcmssm}. The latter condition
sets a relation between the supersymmetry breaking bilinear, trilinear and scalar mass terms.
Because of this condition, $\tan \beta$ is no longer a free parameter, but
must be solved for through the radiative electroweak symmetry breaking relations.

\begin{figure}[hb]
\begin{center}
  \includegraphics[width=0.55\textwidth]{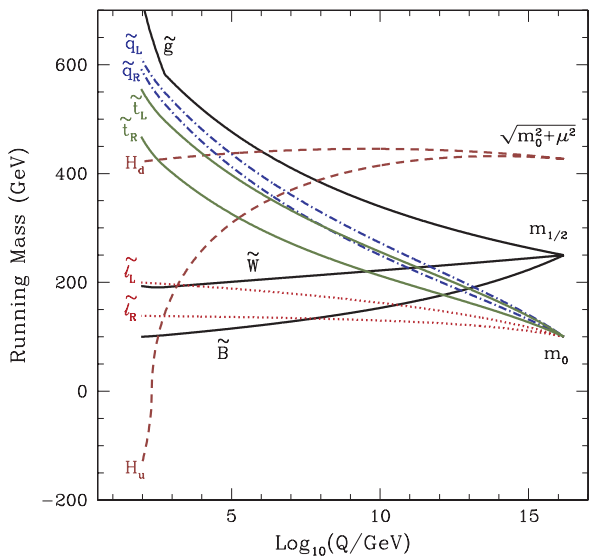}
  \end{center}
  \vskip -.3in
\caption{RG evolution of the mass parameters in the CMSSM.
I thank Toby Falk for providing this figure.}
\label{running}       
\end{figure}

In Fig.~\ref{running}, an example of the renormalization group running of the mass parameters
in the CMSSM is shown.  Here, we have chosen $m_{1/2} = 250$ GeV, $m_0 = 100$
GeV, $\tan \beta = 3$, $A_0 = 0$, and $\mu < 0$.
Indeed, it is rather amazing that from so few input parameters, all of the
masses of the supersymmetric particles can be determined. 
The characteristic features that one sees in the figure, are for example, that
the colored sparticles are typically the heaviest in the spectrum.  This is
due to the large positive correction to the masses due to $\alpha_3$ in the
RGE's.  Also, one finds that the $\widetilde{B}$, is typically the lightest
sparticle.  But most importantly, notice that one of the Higgs mass$^2$, goes
negative triggering electroweak symmetry breaking \cite{rewsb}. (The negative
sign in the figure refers to the sign of the mass squared, even though it is the
mass of the sparticles which is depicted.) 

\begin{figure}[hb]
\begin{center}
  \includegraphics[width=0.45\textwidth]{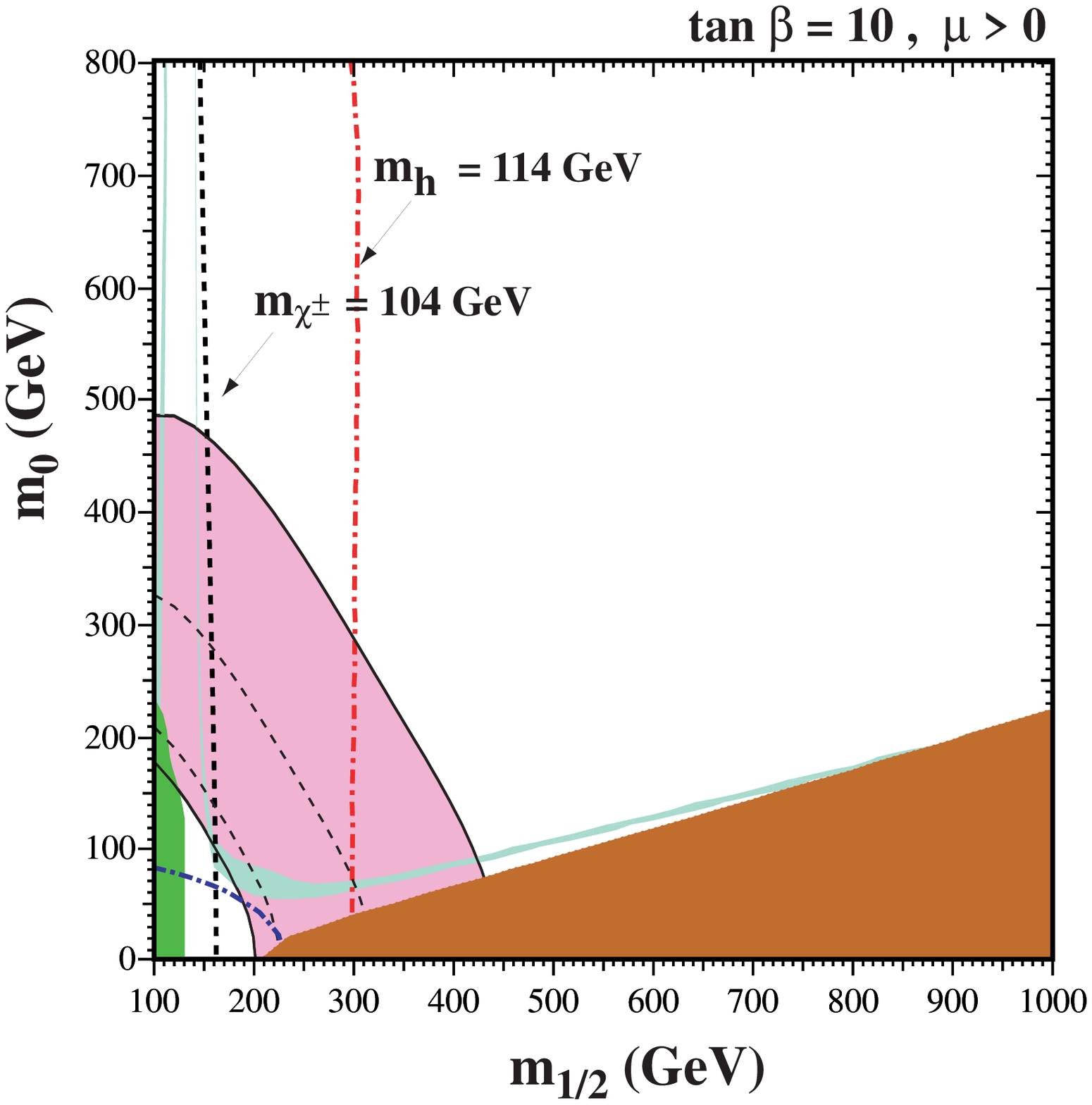}
  \includegraphics[width=0.47\textwidth]{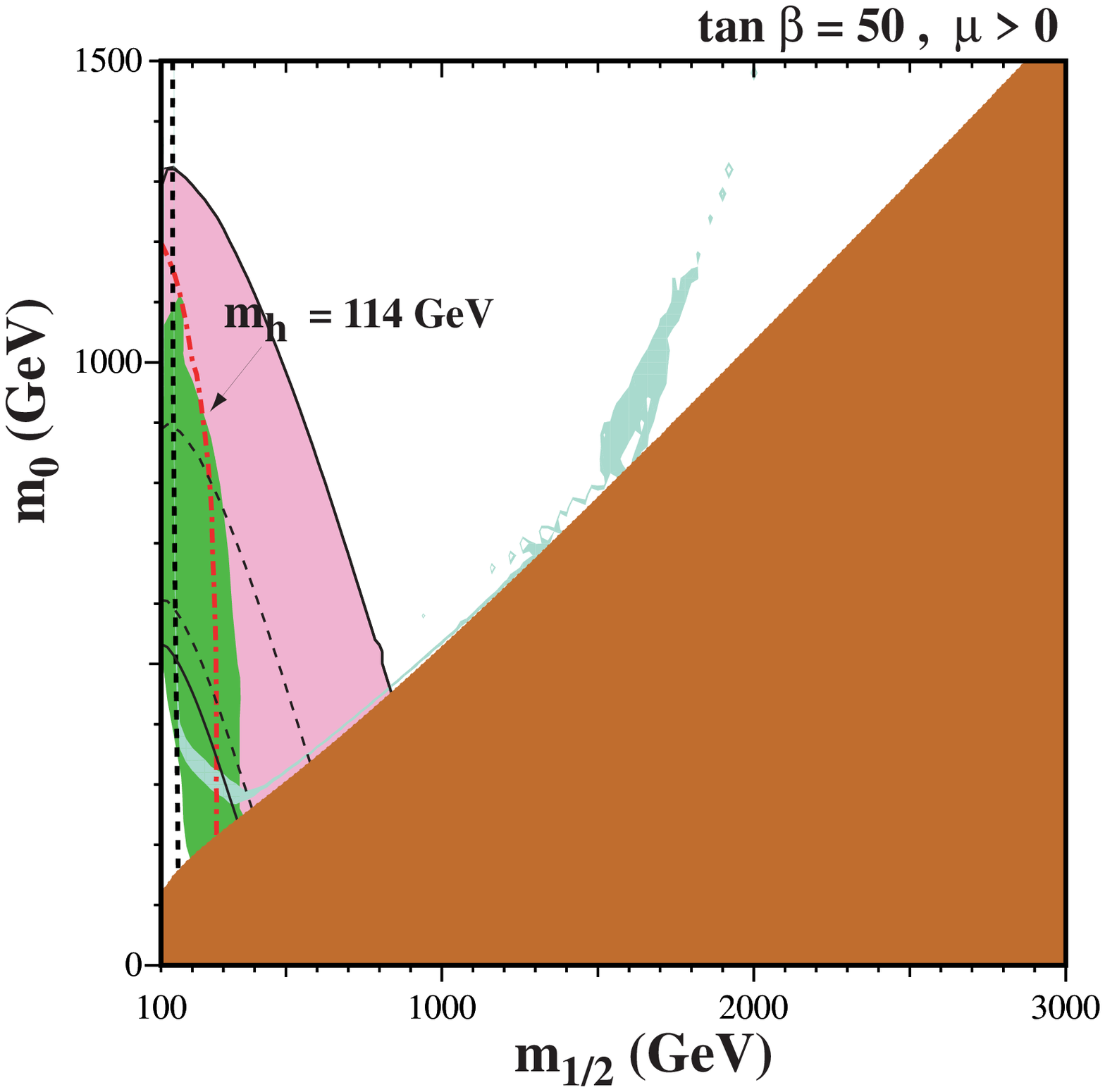}
\end{center}
\caption{The $(m_{1/2}, m_0)$ planes for  (a) $\tan \beta = 10$ and  $\mu > 0$, 
assuming $A_0 = 0, m_t = 175$~GeV and
$m_b(m_b)^{\overline {MS}}_{SM} = 4.25$~GeV. The near-vertical (red)
dot-dashed lines are the contours $m_h = 114$~GeV, and the near-vertical (black) dashed
line is the contour $m_{\chi^\pm} = 104$~GeV. Also
shown by the dot-dashed curve in the lower left is the corner
excluded by the LEP bound of $m_{\tilde e} > 99$ GeV. The medium (dark
green) shaded region is excluded by $b \to s
\gamma$, and the light (turquoise) shaded area is the cosmologically
preferred region. In the dark
(brick red) shaded region, the LSP is the charged ${\tilde \tau}_1$. The
region allowed by the E821 measurement of $a_\mu$ at the 2-$\sigma$
level, is shaded (pink) and bounded by solid black lines, with dashed
lines indicating the 1-$\sigma$ ranges. In (b), $\tan \beta= 50$. }
\label{fig:UHM}       
\end{figure}

For given values of $\tan \beta$, $A_0$,  and $sgn(\mu)$, the regions of the CMSSM
parameter space that yield an
acceptable relic density and satisfy the other phenomenological constraints
may be displayed in the  $(m_{1/2}, m_0)$ plane.
In Fig. \ref{fig:UHM}a,  the light
shaded region corresponds to that portion of the CMSSM plane
with $\tan \beta = 10$, $A_0 = 0$, and $\mu > 0$ such that the computed
relic density yields the WMAP value given in eq. (\ref{wmap}) \cite{eoss}.
The bulk region at relatively low values of 
$m_{1/2}$ and $m_0$,  tapers off
as $m_{1/2}$ is increased.  At higher values of $m_0$,  annihilation cross sections
are too small to maintain an acceptable relic density and $\Omega_\chi h^2$ is too large.
Although sfermion masses are also enhanced at large $m_{1/2}$ (due to RGE running),
co-annihilation processes between the LSP and the next lightest sparticle 
(in this case the $\tilde \tau$) enhance the annihilation cross section and reduce the
relic density.  This occurs when the LSP and NLSP are nearly degenerate in mass.
The dark shaded region has $m_{\tilde \tau}< m_\chi$
and is excluded.   The effect of coannihilations is
to create an allowed band about 25-50 GeV wide in $m_0$ for $m_{1/2} \la
950$ GeV, or $m_{1/2} \la 400$ GeV, which tracks above the $m_{{\tilde \tau}_1}
 =m_\chi$ contour~\cite{efo}.  

Also shown in Fig. \ref{fig:UHM}a are
the relevant phenomenological constraints.  
These include the LEP lower limits on the chargino mass: $m_{\chi^\pm} > 104$~GeV~\cite{LEPsusy}
 and on the Higgs mass: $m_h >
114$~GeV~\cite{LEPHiggs}. 
{\tt FeynHiggs}~\cite{FeynHiggs} is used for the calculation of $m_h$.
The Higgs limit  imposes important constraints,
principally on $m_{1/2}$ and particularly at low $\tan \beta$.
Another constraint is the requirement that
the branching ratio for $b \rightarrow
s \gamma$ be consistent with the experimental measurements~\cite{bsgex}. 
These measurements agree with the Standard Model, and
therefore provide bounds on MSSM particles~\cite{gam},  such as the chargino and
charged Higgs bosons, in particular.  The constraint imposed by
measurements of $b\rightarrow s\gamma$ also exclude small
values of $m_{1/2}$. Finally, there are
regions of the $(m_{1/2}, m_0)$ plane that are favored by
the BNL measurement~\cite{newBNL} of the anomalous magnetic moment of the muon or 
$g_\mu - 2$. Here, we assume
the Standard Model calculation~\cite{Davier} of $g_\mu - 2$,
 and indicate by dashed and solid lines the contours
of 1- and 2-$\sigma$ level deviations induced by supersymmetry.  

At larger $m_{1/2}, m_0$ and $\tan \beta$, the relic neutralino
density may be reduced by rapid annihilation through direct-channel $H, A$ Higgs 
bosons, as seen in Fig.~\ref{fig:UHM}(b) \cite{funnel,efgosi}.
Finally, the relic density can again be brought
down into the WMAP range at large $m_0$ (not shown in 
Fig.~\ref{fig:UHM}), in the `focus-point' region close the boundary where electroweak 
symmetry breaking ceases to be possible and the lightest neutralino $\chi$
acquires a significant higgsino component \cite{fp}.

As seen in Fig.~\ref{fig:UHM}, the relic density constraint is compatible
with relatively large values of $m_{1/2}$ and $m_0$. However, all
values of $m_{1/2}$ and $m_0$ are not equally viable when 
the available phenomenological and cosmological constraints are taken into account.
A global likelihood analysis enables one to 
pin down the available parameter space in the CMSSM. 
One can avoid the dependence on priors by performing
a pure likelihood analysis as in~\cite{Ellis:2003si}, or a purely $\chi^2$-based fit as 
done in~\cite{Ellis:2007fu,Buchmueller:2007zk}.  
Here, the results from one such analysis~\cite{Buchmueller:2008qe,Buchmueller:2009fn} 
is presented, 
which used a Markov-Chain
Monte Carlo (MCMC) technique to explore efficiently the likelihood function in
the parameter space of the CMSSM. A full list of the observables and the values assumed
for them in this global analysis are given in~\cite{Buchmueller:2007zk}, as updated 
in~\cite{Buchmueller:2008qe,Buchmueller:2009fn}. 

\begin{figure}[hb]
\begin{center}
  \includegraphics[width=0.6\textwidth]{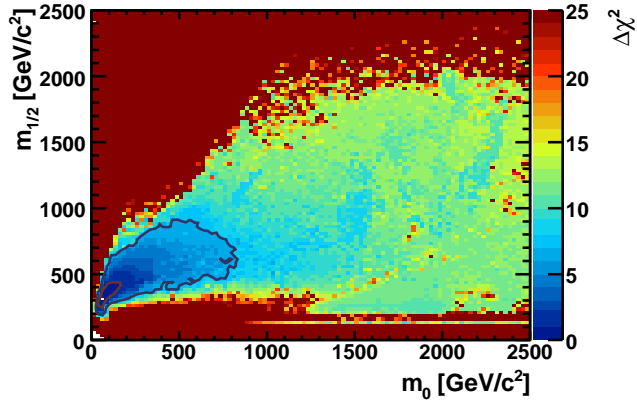}
  \end{center}
\caption{The $\Delta\chi^2$ functions in the $(m_0, m_{1/2})$ planes for
  the CMSSM. We see that the coannihilation region at
  low $m_0$ and $m_{1/2}$ is favored.}
\label{fig:m0m12}       
\end{figure}

The best fit point is shown in  Fig.~\ref{fig:m0m12},
which also displays contours of the $\Delta\chi^2$ function in the CMSSM.
The parameters of the best-fit CMSSM point are
$m_0 = 60 \gev$,  $m_{1/2} = 310 \gev$,  $A_0 = 130 \gev$, $\tb = 11$
and $\mu = 400 \gev$, 
yielding the overall $\chi^2/{\rm N_{\rm dof}} = 20.6/19$ (36\% probability) 
and nominally $\Mh = 114.2 \gev$ \cite{Buchmueller:2009fn}.
The best-fit point is in the coannihilation region of the
$(m_0, m_{1/2})$ plane. The C.L.\ contours extend to slightly larger
values of $m_0$ in the CMSSM.
However, the qualitative features of the $\Delta\chi^2$ contours 
indicate a preference for small
$m_0$ and $m_{1/2}$. It was found in~\cite{Buchmueller:2008qe} that
the focus-point region was disfavored at beyond the 95\% C.L. in 
the CMSSM. We see in Fig.~\ref{fig:m0m12} that this
region is disfavored at the level $\Delta\chi^2 \sim 8$ in the CMSSM. 

As noted above, there are several important cosmological and phenomenological constraints on the supersymmetric parameter space. 
Improvements in sensitivity have made it possible for direct detection experiments \cite{cdms,xenon10}
to be competitive as well.
The elastic cross section for $\chi$ scattering on a nucleus can be
decomposed into a scalar (spin-independent)  and a spin-dependent
part.  Each of these can be
written in terms of the cross sections for elastic scattering off individual nucleons.
The scalar part of the cross section can be
written as
\begin{equation} \label{eqn:sigmaSI}
  \sigma_{\rm SI} = \frac{4 m_{r}^{2}}{\pi}
                    \left[ Z f_{p} + (A-Z) f_{n}  \right]^{2},
\end{equation}
where $m_r$ is the $\chi$-nuclear reduced mass and
\begin{equation} \label{eqn:fN}
  \frac{f_N}{m_N}
    = \sum_{q=u,d,s} \fNTq{q} \frac{\alpha_{3q}}{m_{q}}
      + \frac{2}{27} f_{TG}^{(N)}
        \sum_{q=c,b,t} \frac{\alpha_{3q}}{m_q} ,
\end{equation}
for $N$ = p or n.  The parameters $\fNTq{q}$ are defined by
\begin{equation} \label{eqn:Bq}
  m_N \fNTq{q}
  \equiv \langle N | m_{q} \bar{q} q | N \rangle
  \equiv m_q \BNq{q} ,
\end{equation}
and the $\alpha_{3q}$ contain the individual quark-neutralino scattering cross sections,
see \cite{EFlO1,eoss8,eosv} for further details regarding the calculation of the cross section.

The elastic scattering of neutralinos on nucleons is very sensitive to the strangeness
contribution to the nucleon mass and can be characterized by the parameter, $y$, 
which is also related to the $\pi$-nucleon sigma term $\SigmapiN$ by
\begin{equation} \label{eqn:y2}
  y \equiv {2B_s \over B_u + B_d} = 1 - \sigma_0/\SigmapiN \; .
\end{equation}
where $\sigma_0$ is the change in the nucleon mass due to nonzero $u$ and $d$ masses and is estimated from octet baryon mass differences to be 
$\sigma_0 = 36$~MeV~\cite{Borasoy:1996bx},
and the latest determination of $\SigmapiN = 64$~MeV.
The effect of varying these assumptions are discussed in the context of the CMSSM
in~\cite{eoss8,eosv}. Lattice calculations are now reaching the stage where they may also provide useful information on $\Sigma_{\pi N}$~\cite{0901.3310}, and a recent analysis
would suggest a lower value $\Sigma_{\pi N} \la 40$ \cite{joel}.

\begin{figure}[hbt]
\begin{center}
  \includegraphics[width=0.45\textwidth]{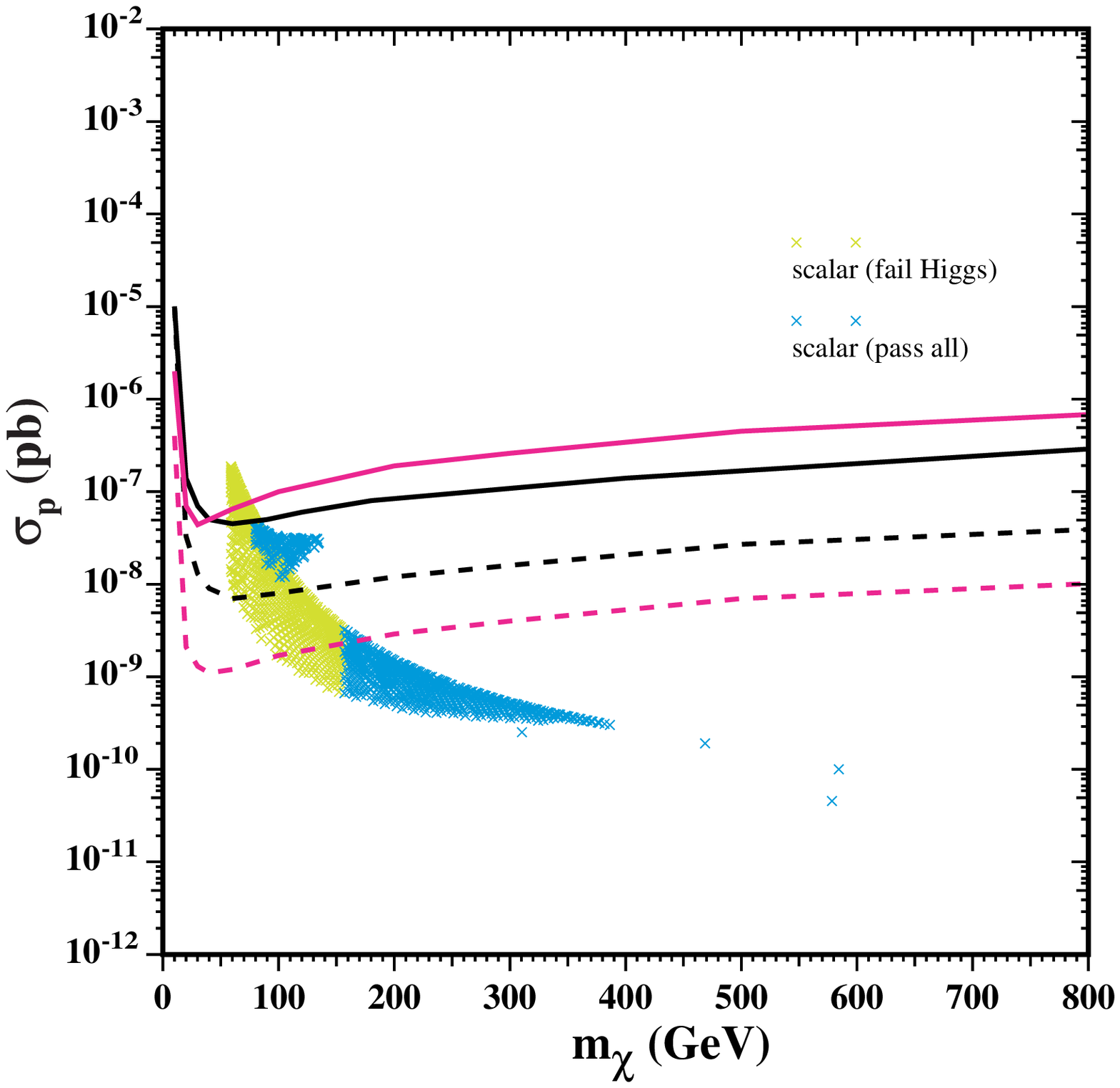}
  \includegraphics[width=0.45\textwidth]{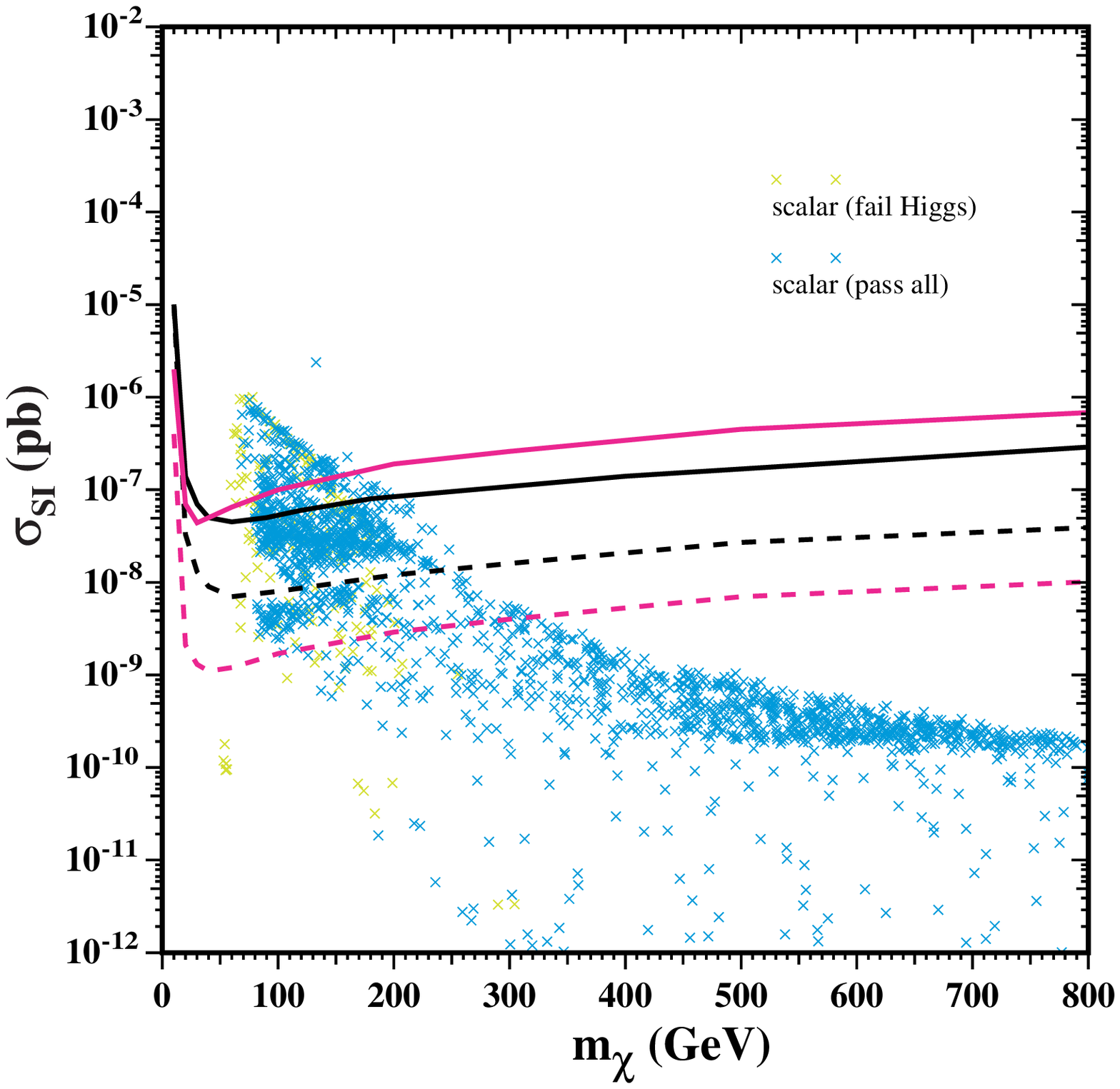}
\end{center}
\caption{The neutralino-nucleon cross sections as functions of neutralino mass for the CMSSM with $ \tan \beta = 10$. Also shown are upper limits on the 
cross section from CDMS~II~\protect\cite{cdms} (solid black line) and
XENON10~\cite{xenon10} (solid pink line), as well as the expected sensitivities  
for XENON100~\cite{XENON100} (dashed pink line) and SuperCDMS at the Soudan 
Mine~\protect\cite{superCDMS} (dashed black line). 
Panel (b) shows the neutralino-nucleon cross sections as functions of neutralino mass for the CMSSM, with $5 \leq \tan \beta \leq 55$, 0 $\leq m_{1/2} \leq 2000$ GeV, 100 GeV $\leq m_0 \leq 2000$ GeV, and $-3 m_{1/2} \leq A_0 \leq 3 m_{1/2}$  We consider $\mu<0$ only for $\tan \beta<30$. Taken from \protect\cite{eosk5}.
\label{fig:cmssm}}
\end{figure}

In panel (a) of Fig.~\ref{fig:cmssm} the spin-independent 
neutralino-nucleon elastic scattering cross sections are shown as
functions of neutralino mass for the regions of Fig.~\ref{fig:UHM}a that are cosmologically viable 
(i.e., those where the upper limit on the relic density of neutralinos is respected), 
and are not excluded by constraints from colliders.  Here, however, parameter values corresponding 
to the focus point at high $m_0$ are also included.
Also plotted are the limits on the 
spin-independent cross section from CDMS~II~\cite{cdms} (solid black line)
and XENON10~\cite{xenon10} (solid red line), as well as the sensitivities projected  
for XENON100~\cite{XENON100} (or a similar 100-kg liquid noble-gas detector such as LUX, 
dashed red line) and SuperCDMS at the Soudan Mine~\cite{superCDMS} (dashed black line).

There are two distinct 
regions in the $(m_{\chi},\sigma)$ plane, that arising from the focus-point region
at $m_{\chi} \lesssim 150$ GeV and relatively large $\sigma$, and that from the 
coannihilation strip.  In the coannihilation strip, 50 GeV $< m_{\chi} <$ 400 GeV, 
where the lower limit on $m_{\chi}$ is a result of the LEP constraint on the chargino mass,
and the upper limit on $m_{\chi}$ corresponds to the end-point of the coannihilation strip 
for $\tan \beta = 10$.  In contrast, the end point of the focus-point region shown is due 
only to the cut-off $m_0 < 2$ TeV that has been assumed.  
In addition, for $m_{1/2} \lesssim 380$~GeV 
in the coannihilation strip ($m_{\chi} \lesssim 160$ GeV), the nominal calculated mass of the lighter
scalar MSSM Higgs boson is less than the LEP lower bound.  These points are indicated
by lighter shadings.

The choices $\tb=10$ and $A_0=0$ do not yield viable direct detection 
cross sections that are completely representative of the range of possibilities within the CMSSM.
Therefore, in Fig.~\ref{fig:cmssm}b, we show CMSSM spin-independent
neutralino-nucleon cross section,  as obtained in a scan over all 
CMSSM parameters with $5 \leq \tan \beta \leq 55$, 100 $\leq m_{1/2} \leq 2000$ GeV, 
0 GeV $\leq m_0 \leq 2000$ GeV, and $-3 m_{1/2} \leq A_0 \leq 3 m_{1/2}$ \cite{eosk5}.  
We also allow both positive and negative $\mu$, except for large $\tan \beta > 30$, 
where convergence becomes difficult in the $\mu < 0$ case.  
 At low $m_{\chi} < 300$~GeV, cross sections 
generally exceed $10^{-9}$~pb, and the largest scalar cross sections, 
which occur for $m_{\chi} \sim 100$ GeV, are already excluded by CDMS~II \cite{cdms} and/or 
XENON10 \cite{xenon10}. 
These exclusions occur primarily in the focus-point region at large $\tan \beta$.  
On the other hand, for $m_{\chi} \ga 400$ GeV scalar cross sections are well 
below $10^{-9}$ pb, and come from the coannihilation strip or the rapid-annihilation 
funnel that appears at large $\tan \beta$ in the CMSSM. The effective cross sections shown are
suppressed for points with $\Omega_\chi \ll \Omega_{CDM}$, and there may be cancellations
at larger $m_{\chi}$ that suppress the cross sections substantially. 
These regions of parameter space will not be probed by direct detection experiments in the near future
\cite{XENON100,superCDMS}.  

\begin{figure}[hbt]
\begin{center}
  \includegraphics[width=0.45\textwidth]{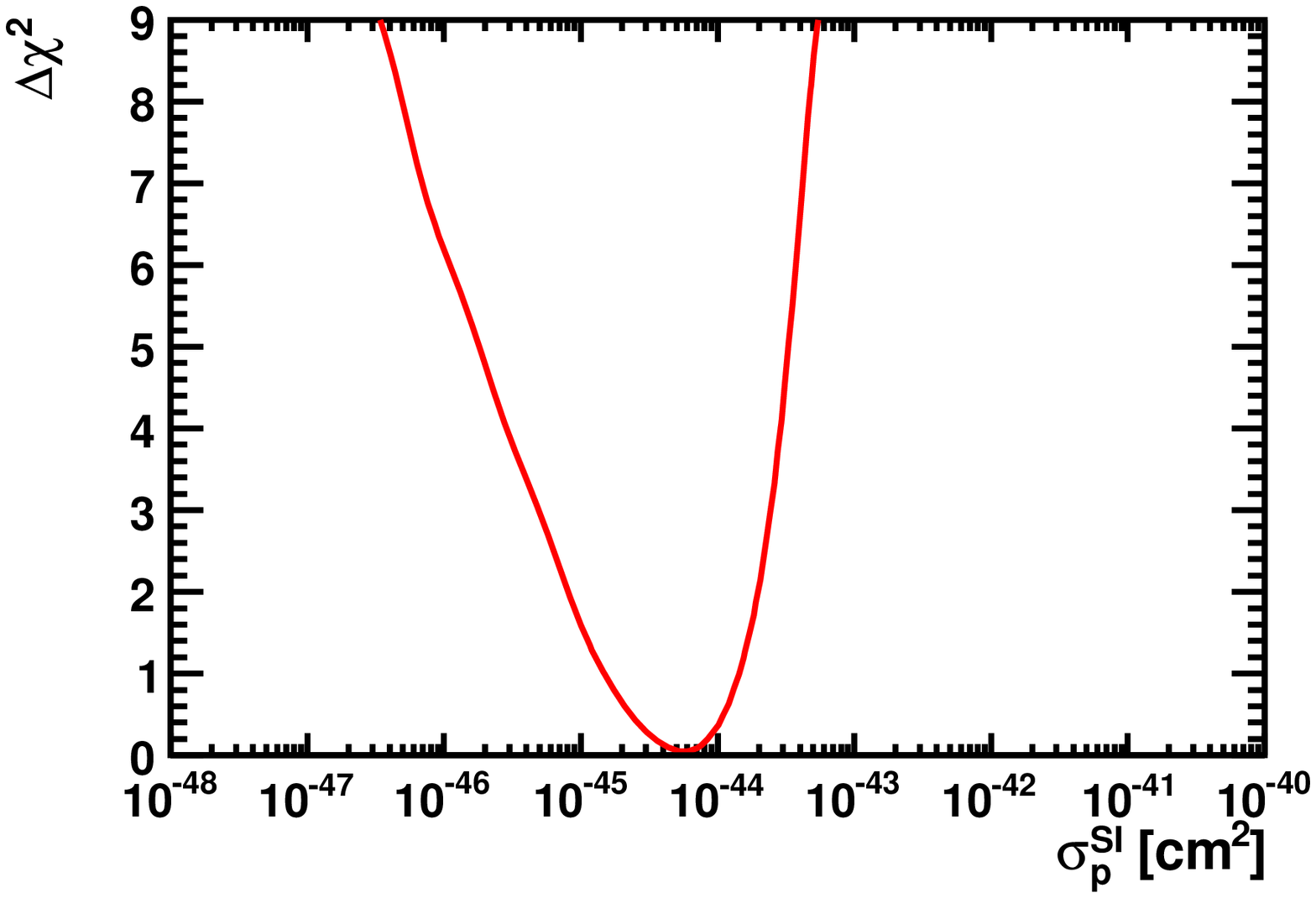}
  \includegraphics[width=0.45\textwidth]{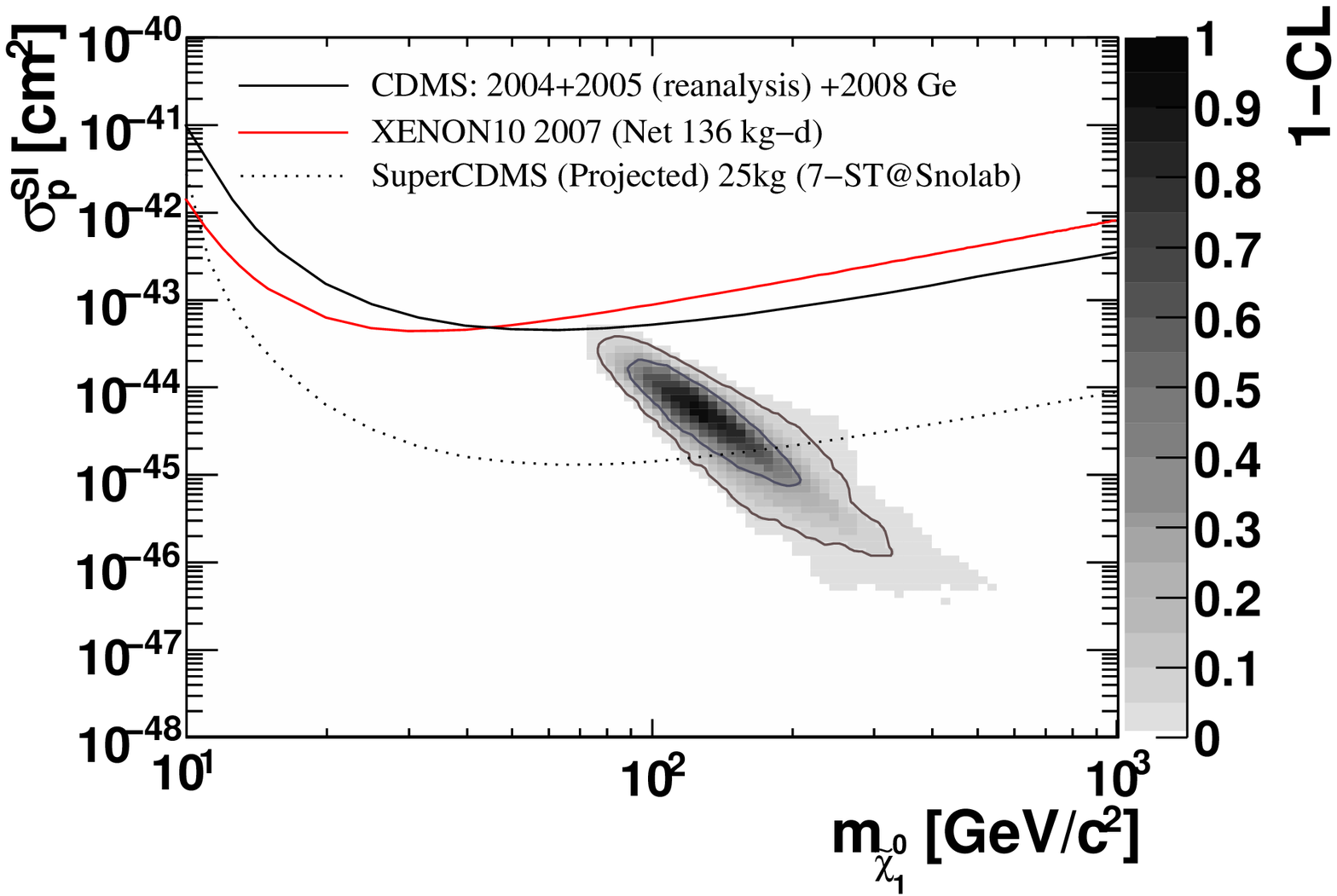}
\end{center}
\caption{The likelihood functions for the
  spin-independent $\neu{1}$-proton scattering cross section $\ssi$ (in cm$^2$)
in the CMSSM (left panel). The correlation between the spin-independent DM
scattering cross section $\ssi$ and $\mneu{1}$
in the CMSSM (right panel).
\label{fig:sig}}
\end{figure}

Finally, the frequentist analysis described above can also be used to predict
the neutralino-nucleon elastic scattering cross section \cite{Buchmueller:2009fn}. 
The value of $\ssi$ shown in Fig.~\ref{fig:sig}a is calculated assuming a
  $\pi$-N scattering $\sigma$ term 
$\Sigma_N = 64$~MeV.
We see in Fig.~\ref{fig:sig} that values of the $\neu{1}$-proton cross
section $\ssi \sim 10^{-8}$~pb are expected in the CMSSM, and that much
larger values seem quite unlikely. The 2D $\chi^2$ function is shown in Fig. ~\ref{fig:sig}b.

\section{Acknowledgments}

This work was supported in part by DOE grant DE-FG02-94ER-40823.


\begin{footnotesize}




\begin{thebibliography}{99}
\bibitem{wmap} 
  J.~Dunkley {\it et al.}  [WMAP Collaboration],
  Astrophys.\ J.\ Suppl.\  {\bf 180}, 306 (2009)
  [arXiv:0803.0586 [astro-ph]];
  E.~Komatsu {\it et al.}  [WMAP Collaboration],
  Astrophys.\ J.\ Suppl.\  {\bf 180}, 330 (2009)
  [arXiv:0803.0547 [astro-ph]].
  
   \bibitem{sn1}
  A.G. Riess \etal, A.\ J.\ {\bf 116}, 1009 (1998);
  P. Garnavich \etal, Astrophys.\ J.\ 
        {\bf 509}, 74
        (1998);
        S. Perlmutter \etal, Ap.\ J.\ {\bf 517}, 565 (1999);
  A. G. Riess {\it et al.}, {\it Ap.~J.} {\bf 560} (2001) 49;
  J.L. Tonry \etal, Ap.\ J.\ {\bf 594}, 1 (2003);
  A.G. Riess \etal, Astrophys.\ J.\ {\bf 659}, 98 (2007);
     P.~Astier {\it et al.}  [The SNLS Collaboration],
  Astron.\ Astrophys.\  {\bf 447}, 31 (2006)
  [arXiv:astro-ph/0510447].


\bibitem{bao}
 D. Eisenstein  \etal, Astrophys.\ J.\
{\bf 633}, 560 (2005);
C. Blake  \etal, Mon.\ Not.\ Roy.\ Astr.\ Soc.\, 
{\bf 374}, 1527 (2007);
 W.~J.~Percival, \etal,
  Mon.\ Not.\ Roy.\ Astron.\ Soc.\  {\bf 381}, 1053 (2007);
 M. Kowalski  \etal, Astrophys.\ J.\ {\bf 686}, 749 (2008).

\bibitem{cc}
  S.~Weinberg,
  Rev.\ Mod.\ Phys.\  {\bf 61}, 1 (1989);
  S.~M.~Carroll,
  Living Rev.\ Rel.\  {\bf 4}, 1 (2001)
  [arXiv:astro-ph/0004075];
    P.~J.~E.~Peebles and B.~Ratra,
  Rev.\ Mod.\ Phys.\  {\bf 75}, 559 (2003)
  [arXiv:astro-ph/0207347].

\bibitem{quint}
 I.~Zlatev, L.~M.~Wang and P.~J.~Steinhardt,
  Phys.\ Rev.\ Lett.\  {\bf 82}, 896 (1999)
  [arXiv:astro-ph/9807002];
  C.~Armendariz-Picon, V.~F.~Mukhanov and P.~J.~Steinhardt,
  Phys.\ Rev.\  D {\bf 63}, 103510 (2001)
  [arXiv:astro-ph/0006373].

\bibitem{rot}
S. M. Faber and J. J. Gallagher,  {\it Ann. Rev. Astron.
Astrophys.} {\bf 17} (1979) 135;
A. Bosma, {\it Ap. J.} {\bf 86} (1981) 1825; V. C. Rubin, W. K.
Ford and N. Thonnard, {\it Ap. J.} {\bf 238} (1980) 471;	V. C. Rubin, D.
Burstein, W. K. Ford and N. Thonnard, {\it Ap. J.} {\bf 289} (1985) 81;
T. S. Van Albada  and R. Sancisi, {\it Phil. Trans. R. Soc. Land.} {\bf
A320} (1986) 447;
M. Persic and P. Salucci, {\it Ap. J. Supp.} {\bf 99} (1995) 501;
M. Persic, P. Salucci, and F. Stel, {\it MNRAS} {\bf 281} (1996) 27P.

  
   \bibitem{witt} 
 D.~Wittman, {\it et al.},
  Astrophys.\ J.\  {\bf 643}, 128 (2006)
  [arXiv:astro-ph/0507606].

\bibitem{clowe}
D.~Clowe,{\it et al.},
   Astrophys.\ J.\  {\bf 648} (2006) L109
[arXiv:astro-ph/0608407].
  
    \bibitem{otasi3}
K.~A.~Olive,
arXiv:astro-ph/0301505.

\bibitem{isotopes} J. Rich, M. Spiro and J. Lloyd-Owen, {\it Phys.Rep.}
{\bf 151} (1987) 239;
  T.~Yamagata, Y.~Takamori and H.~Utsunomiya,
  Phys.\ Rev.\  D {\bf 47} (1993) 1231;
  T. K. Hemmick {\it et al.}, Phys. Rev.  {\bf D41} (1990) 2074;
P. F. Smith, {Contemp. Phys.} {\bf 29} (1998) 159;
 D.~Javorsek, D.~Elmore, E.~Fischbach, D.~Granger, T.~Miller, D.~Oliver and V.~Teplitz,
  Phys.\ Rev.\  D {\bf 65}, 072003 (2002).

\bibitem{mssm}
P.~Fayet,
{Phys. Lett. }{\bf B64} (1976) {159};
{Phys. Lett. }{\bf B69} (1977) {489};
{Phys. Lett. }{\bf B84} (1979) {416};
H.E. Haber and G.L. Kane, {\it Phys.Rep.} {\bf 117} (1985) 75.

\bibitem{Falk:1994es}
T.~Falk, K.~A.~Olive and M.~Srednicki,
  Phys.\ Lett.\  B {\bf 339}, 248 (1994)
  [arXiv:hep-ph/9409270];
  C.~Arina and N.~Fornengo,
  JHEP {\bf 0711}, 029 (2007)
  [arXiv:0709.4477 [hep-ph]].
  
\bibitem{LEP:2003ih}
LEP Collaborations,  ALEPH, DELPHI, L3 and OPAL, LEP Electroweak Working Group, SLD Electroweak Group and SLD Heavy Flavor Group, hep-ex/0312023.

\bibitem{vcmssm}
J.~R.~Ellis, K.~A.~Olive, Y.~Santoso and V.~C.~Spanos,
  Phys.\ Lett.\ B {\bf 573} (2003) 162
  [arXiv:hep-ph/0305212];
  J.~R.~Ellis, K.~A.~Olive, Y.~Santoso and V.~C.~Spanos,
  Phys.\ Rev.\ D {\bf 70} (2004) 055005
  [arXiv:hep-ph/0405110].

 \bibitem{gdm}
  J.~R.~Ellis, K.~A.~Olive, Y.~Santoso and V.~C.~Spanos,
  Phys.\ Lett.\ B {\bf 588} (2004) 7
  [arXiv:hep-ph/0312262];
J.~L.~Feng, A.~Rajaraman and F.~Takayama,
  Phys.\ Rev.\ Lett.\  {\bf 91} (2003) 011302
  [arXiv:hep-ph/0302215];
  J.~L.~Feng, S.~Su and F.~Takayama,
  Phys.\ Rev.\ D {\bf 70} (2004) 075019
  [arXiv:hep-ph/0404231].

  \bibitem{EHNOS}
J. Ellis, J.S. Hagelin, D.V. Nanopoulos, K.A. Olive
and M. Srednicki, Nucl. Phys. B {\bf 238} (1984) 453; see also
H. Goldberg, Phys. Rev. Lett. {\bf 50} (1983) 1419.

\bibitem{GUT}
J.~R.~Ellis, S.~Kelley and D.~V.~Nanopoulos,
  Phys.\ Lett.\ B {\bf 249} (1990) 441 and
  Phys.\ Lett.\ B {\bf 260} (1991) 131;
U.~Amaldi, W.~de Boer and H.~Furstenau,
  Phys.\ Lett.\ B {\bf 260} (1991) 447;
  C.~Giunti, C.~W.~Kim and U.~W.~Lee,
  Mod.\ Phys.\ Lett.\ A {\bf 6} (1991) 1745.

\bibitem{BIM}
For reviews, see:
H.~P.~Nilles, Phys. Rep. {\bf 110} (1984) 1;
A.~Brignole, L.~E.~Ibanez and C.~Munoz,
arXiv:hep-ph/9707209,
published in {\it Perspectives on supersymmetry}, ed.
G.~L.~Kane, pp. 125-148. 

\bibitem{funnel}
M.~Drees and M.~M.~Nojiri,
Phys.\ Rev.\ D {\bf 47} (1993) 376 [arXiv:hep-ph/9207234];
H.~Baer and M.~Brhlik,
Phys.\ Rev.\ D {\bf 53} (1996) 597 [arXiv:hep-ph/9508321];
  Phys.\ Rev.\  D {\bf 57} (1998) 567
  [arXiv:hep-ph/9706509];
H.~Baer, M.~Brhlik, M.~A.~Diaz, J.~Ferrandis, P.~Mercadante, P.~Quintana and X.~Tata,
  Phys.\ Rev.\  D {\bf 63} (2001) 015007
  [arXiv:hep-ph/0005027];
 A.~B.~Lahanas, D.~V.~Nanopoulos and V.~C.~Spanos,
  Mod.\ Phys.\ Lett.\  A {\bf 16} (2001) 1229
  [arXiv:hep-ph/0009065].

\bibitem{cmssm}
J.~R.~Ellis, T.~Falk, K.~A.~Olive and M.~Schmitt,
Phys.\ Lett.\ B {\bf 388} (1996) 97
[arXiv:hep-ph/9607292];
Phys.\ Lett.\ B {\bf 413} (1997) 355
[arXiv:hep-ph/9705444];
J.~R.~Ellis, T.~Falk, G.~Ganis, K.~A.~Olive and M.~Schmitt,
Phys.\ Rev.\ D {\bf 58} (1998) 095002
[arXiv:hep-ph/9801445];
V.~D.~Barger and C.~Kao,
Phys.\ Rev.\ D {\bf 57} (1998) 3131
[arXiv:hep-ph/9704403];
J.~R.~Ellis, T.~Falk, G.~Ganis and K.~A.~Olive,
Phys.\ Rev.\ D {\bf 62} (2000) 075010
[arXiv:hep-ph/0004169];
V.~D.~Barger and C.~Kao,
Phys.\ Lett.\ B {\bf 518} (2001) 117
[arXiv:hep-ph/0106189];
L.~Roszkowski, R.~Ruiz de Austri and T.~Nihei,
JHEP {\bf 0108} (2001) 024
[arXiv:hep-ph/0106334];
A.~B.~Lahanas and V.~C.~Spanos,
Eur.\ Phys.\ J.\ C {\bf 23} (2002) 185
[arXiv:hep-ph/0106345];
A.~Djouadi, M.~Drees and J.~L.~Kneur,
JHEP {\bf 0108} (2001) 055
[arXiv:hep-ph/0107316];
U.~Chattopadhyay, A.~Corsetti and P.~Nath,
Phys.\ Rev.\ D {\bf 66} (2002) 035003
[arXiv:hep-ph/0201001];
J.~R.~Ellis, K.~A.~Olive and Y.~Santoso,
New Jour.\ Phys.\  {\bf 4} (2002) 32
[arXiv:hep-ph/0202110];
H.~Baer, C.~Balazs, A.~Belyaev, J.~K.~Mizukoshi, X.~Tata and Y.~Wang,
JHEP {\bf 0207} (2002) 050
[arXiv:hep-ph/0205325];
R.~Arnowitt and B.~Dutta,
arXiv:hep-ph/0211417.

\bibitem{efgosi}
J.~R.~Ellis, T.~Falk, G.~Ganis, K.~A.~Olive and M.~Srednicki,
Phys.\ Lett.\ B {\bf 510} (2001) 236
[arXiv:hep-ph/0102098].


\bibitem{eoss}
J.~R.~Ellis, K.~A.~Olive, Y.~Santoso and V.~C.~Spanos,
Phys.\ Lett.\ B {\bf 565} (2003) 176
[arXiv:hep-ph/0303043].



\bibitem{cmssmwmap}
H.~Baer and C.~Balazs,
  JCAP {\bf 0305}, 006 (2003)
  [arXiv:hep-ph/0303114];
A.~B.~Lahanas and D.~V.~Nanopoulos,
  Phys.\ Lett.\  B {\bf 568}, 55 (2003)
  [arXiv:hep-ph/0303130];
U.~Chattopadhyay, A.~Corsetti and P.~Nath,
  Phys.\ Rev.\  D {\bf 68}, 035005 (2003)
  [arXiv:hep-ph/0303201];
   .~Munoz,
  Int.\ J.\ Mod.\ Phys.\  A {\bf 19}, 3093 (2004)
  [arXiv:hep-ph/0309346].
  
  \bibitem{bfs}
R. Barbieri, S. Ferrara and C.A. Savoy, Phys.\ Lett. {\bf 119B} (1982) 343.

  
\bibitem{rewsb}
L.E.~Ib\'a\~nez and G.G.~Ross, {Phys. Lett. }{\bf B110} (1982) {215}; \\
L.E.~Ib\'a\~nez, {Phys. Lett. }{\bf B118} (1982) {73}; \\
J.~Ellis, D.V.~Nanopoulos and K.~Tamvakis,
{Phys. Lett. }{\bf B121} (1983) {123}; \\
J.~Ellis, J. Hagelin, D.V.~Nanopoulos and K.~Tamvakis,
{Phys. Lett. }{\bf B125} (1983) {275}; \\
L.~Alvarez-Gaum\'e, J.~Polchinski, and M.~Wise,
Nucl. Phys. {\bf B221} (1983) {495}.


\bibitem{efo} J. Ellis, T. Falk, and K.A. Olive, Phys. Lett.  {\bf B444} (1998) 367
[arXiv:hep-ph/9810360];
J. Ellis, T. Falk, K.A. Olive, and M. Srednicki, {\it Astr. Part. Phys.}
{\bf 13} (2000) 181
[Erratum-ibid.\  {\bf 15} (2001) 413]
[arXiv:hep-ph/9905481].

\bibitem{LEPsusy}
  Joint LEP~2 Supersymmetry Working Group,
  {\it Combined LEP Chargino Results, up to 208 GeV}, \\
  {\tt http://lepsusy.web.cern.ch/lepsusy/www/inos{\_}
  moriond01/charginos{\_}pub.html}.
  
 
\bibitem{LEPHiggs}
R.~Barate {\it et al.}  [ALEPH, DELPHI, L3, OPAL Collaborations:
  the LEP Working Group for Higgs boson searches],
  Phys.\ Lett.\  B {\bf 565}, 61 (2003)
  [arXiv:hep-ex/0306033];
  D.~Zer-Zion,
  {\it Prepared for 32nd International Conference on High-Energy
       Physics (ICHEP 04),
       Beijing, China, 16-22 Aug 2004};
  LHWG-NOTE-2004-01, ALEPH-2004-008, DELPHI-2004-042, L3-NOTE-2820,
  OPAL-TN-744,
  {\tt http://lephiggs.web.cern.ch/LEPHIGGS/papers/August2004{\_}MSSM/index.html}.
  
\bibitem{FeynHiggs}
S.~Heinemeyer, W.~Hollik and G.~Weiglein,
{\it Comput.\ Phys.\ Commun.\ } {\bf 124} (2000) 76 
[arXiv:hep-ph/9812320];
S.~Heinemeyer, W.~Hollik and G.~Weiglein,
{\it Eur.\ Phys.\ J.\ C} {\bf 9} (1999) 343 
[arXiv:hep-ph/9812472];
G.~Degrassi, S.~Heinemeyer, W.~Hollik, P.~Slavich and G.~Weiglein,
  Eur.\ Phys.\ J.\ C {\bf 28} (2003) 133
  [arXiv:hep-ph/0212020].;
  M.~Frank, T.~Hahn, S.~Heinemeyer, W.~Hollik, H.~Rzehak and G.~Weiglein,
  JHEP {\bf 0702} (2007) 047
  [arXiv:hep-ph/0611326];
  \texttt{http://www.feynhiggs.de/}

\bibitem{bsgex}
S.~Chen {\it et al.}  [CLEO Collaboration],
Phys.\ Rev.\ Lett.\  {\bf 87} (2001) 251807
[arXiv:hep-ex/0108032];
P.~Koppenburg {\it et al.}  [Belle Collaboration],
Phys.\ Rev.\ Lett.\  {\bf 93} (2004) 061803
[arXiv:hep-ex/0403004].
B.~Aubert {\it et al.}  [BaBar Collaboration],
arXiv:hep-ex/0207076;
E.~Barberio {\it et al.}  [Heavy Flavor Averaging Group (HFAG)],
  arXiv:hep-ex/0603003.

\bibitem{gam}
C. Degrassi, P. Gambino and G.~F. Giudice,
{\it JHEP} {\bf 0012} (2000) 009 [arXiv:hep-ph/0009337],
as implemented by P. Gambino and G. Ganis.

\bibitem{newBNL} [The Muon g-2 Collaboration],
                 {\it Phys. Rev. Lett.} {\bf 92} (2004) 161802, 
                 hep-ex/0401008;
                 G.~Bennett et al.\ [The Muon g-2 Collaboration],
                  {\em Phys.\ Rev.} {\bf D 73} (2006) 072003
                  [arXiv:hep-ex/0602035].

\bibitem{Davier}
M.~Davier, S.~Eidelman, A.~H\"ocker and Z.~Zhang,
               {\it Eur.\ Phys.\ J.}\  {\bf C 31} (2003) 503,
               hep-ph/0308213;
                see also
  M.~Knecht,
  Lect.\ Notes Phys.\  {\bf 629}, 37 (2004)
  [arXiv:hep-ph/0307239];
K.~Melnikov and A.~Vainshtein,
Phys.\ Rev.\  {\bf D70} (2004) 113006
[arXiv:hep-ph/0312226];
J.~F.~de Troconiz and F.~J.~Yndurain,
  Phys.\ Rev.\  D {\bf 71}, 073008 (2005)
  [arXiv:hep-ph/0402285];
   M.~Passera,
  arXiv:hep-ph/0411168;
  K.~Hagiwara, A.~Martin, D.~Nomura and T.~Teubner,
                   {\em Phys.\ Lett.} {\bf B 649} (2007) 173
                   [arXiv:hep-ph/0611102];
  M.~Davier,
  Nucl.\ Phys.\ Proc.\ Suppl.\  {\bf 169}, 288 (2007)
  [arXiv:hep-ph/0701163];
F.~Jegerlehner,
  Acta Phys.\ Polon.\  B {\bf 38}, 3021 (2007)
  [arXiv:hep-ph/0703125];
   J.~Miller, E.~de~Rafael and B.~Roberts,
                       {\em Rept.\ Prog.\ Phys.} {\bf 70} (2007) 795
                       [arXiv:hep-ph/0703049];
S.~Eidelman, 
            talk given at the ICHEP06, Moscow, July 2006, see:\\
            {\tt http://ichep06.jinr.ru/reports/333\_6s1\_9p30\_
            Eidelman.pdf};
  M.~Davier, A.~Hoecker, B.~Malaescu, C.~Z.~Yuan and Z.~Zhang,
  arXiv:0908.4300 [hep-ph].

\bibitem{fp}
J.~L.~Feng, K.~T.~Matchev and T.~Moroi,
{\it Phys.\ Rev.\ D} {\bf 61} (2000) 075005
[arXiv:hep-ph/9909334].

\bibitem{Ellis:2003si}
  J.~R.~Ellis, K.~A.~Olive, Y.~Santoso and V.~C.~Spanos,
  Phys.\ Rev.\  D {\bf 69} (2004) 095004
  [arXiv:hep-ph/0310356].

\bibitem{Ellis:2007fu}
  J.~R.~Ellis, S.~Heinemeyer, K.~A.~Olive, A.~M.~Weber and G.~Weiglein,
  JHEP {\bf 0708}, 083 (2007)
  [arXiv:0706.0652 [hep-ph]].

\bibitem{Buchmueller:2007zk} O.~Buchmueller {\it et al.},
  Phys.\ Lett.\  B {\bf 657} (2007) 87
  [arXiv:0707.3447 [hep-ph]].

 \bibitem{Buchmueller:2008qe} O.~Buchmueller {\it et al.},
  JHEP {\bf 0809} (2008) 117
  [arXiv:0808.4128 [hep-ph]].
        
  \bibitem{Buchmueller:2009fn}
  O.~Buchmueller {\it et al.},
  Eur.\ Phys.\ J.\  C {\bf 64}, 391 (2009)
  [arXiv:0907.5568 [hep-ph]].

\bibitem{cdms}
  Z.~Ahmed {\it et al.}  [CDMS Collaboration],
  Phys.\ Rev.\ Lett.\  {\bf 102}, 011301 (2009)
  [arXiv:0802.3530 [astro-ph]].

\bibitem{xenon10}
  J.~Angle {\it et al.}  [XENON Collaboration],
  Phys.\ Rev.\ Lett.\  {\bf 100}, 021303 (2008)
  [arXiv:0706.0039 [astro-ph]].

  \bibitem{EFlO1}
J.~Ellis, A.~Ferstl and K.~A.~Olive,
Phys. Lett. B {\bf 481}, 304 (2000)
[arXiv:hep-ph/0001005];
J.~Ellis, A.~Ferstl and K.~A.~Olive,
Phys.\ Rev.\ D {\bf 63}, 065016 (2001)
[arXiv:hep-ph/0007113];
J.~R.~Ellis, A.~Ferstl and K.~A.~Olive,
Phys.\ Lett.\ B {\bf 532}, 318 (2002)
[arXiv:hep-ph/0111064].


  \bibitem{eoss8}
  J.~R.~Ellis, K.~A.~Olive, Y.~Santoso and V.~C.~Spanos,
  Phys.\ Rev.\  D {\bf 71}, 095007 (2005)
  [arXiv:hep-ph/0502001].
  
  \bibitem{eosv}
  J.~R.~Ellis, K.~A.~Olive and C.~Savage,
  Phys.\ Rev.\  D {\bf 77}, 065026 (2008)
  [arXiv:0801.3656 [hep-ph]].

\bibitem{Borasoy:1996bx}
  B.~Borasoy and U.~G.~Meissner,
  Annals Phys.\  {\bf 254}, 192 (1997)
  [arXiv:hep-ph/9607432];
  J.~Gasser, H.~Leutwyler and M.~E.~Sainio,
  Phys.\ Lett.\  B {\bf 253}, 252 (1991);
  M.~Knecht,
  PiN Newslett.\  {\bf 15}, 108 (1999)
  [arXiv:hep-ph/9912443];
  M.~E.~Sainio,
  PiN Newslett.\  {\bf 16}, 138 (2002)
  [arXiv:hep-ph/0110413].
  
  \bibitem{0901.3310}
R.~D.~Young and A.~W.~Thomas,
  arXiv:0901.3310 [hep-lat].

\bibitem{joel}
  J.~Giedt, A.~W.~Thomas and R.~D.~Young,
  arXiv:0907.4177 [hep-ph].

  \bibitem{eosk5}
  J.~Ellis, K.~A.~Olive and P.~Sandick,
  arXiv:0905.0107 [hep-ph].
  
 \bibitem{XENON100}
  E.~Aprile, L.~Baudis and f.~t.~X.~Collaboration,
  arXiv:0902.4253 [astro-ph.IM].


\bibitem{superCDMS}
  {\it SuperCDMS Development Project},
  Fermilab Proposal 0947,  
  October 2004.

\end{thebibliography}
%

\end{footnotesize}


\end{document}